\documentclass[aps,prd,reprint,preprintnumbers]{revtex4-2}

\usepackage[bookmarks=true]{hyperref}
\usepackage{graphicx,epsfig,color}
\usepackage[english]{babel}
\usepackage{amssymb,amsfonts,amsmath}
\usepackage{verbatim}

\newcommand{\inlinesection}[1]{\textit{#1}.---}

\newcommand{\OO}{\mathcal{O}}

\begin{document}

\title{Black brane evaporation through D-brane bubble nucleation}

\author{Oscar Henriksson}
\email{oscar.henriksson@helsinki.fi}
\affiliation{Department of Physics and Helsinki Institute of Physics\\
P.O.~Box 64, FI-00014 University of Helsinki, Finland}

\begin{abstract}
We study the process of black brane evaporation through the emission of D-branes. Black branes in asymptotically anti--de Sitter spacetimes, holographically dual to field theory states at finite temperature and density, have previously been found to exhibit an instability due to brane nucleation. Working in the setting of D3-branes on the conifold, we construct static Euclidean solutions describing this nucleation to leading order---D3-branes bubbling off the horizon. Furthermore, we analyze the late-time dynamics of such a D3-brane bubble as it expands and find a steady-state solution including the wall profile and its speed.
\end{abstract}

\preprint{HIP-2021-20/TH}

\maketitle

%%%%%%%%%%%%%%%%%%%%%%%%%%%%
\section{Introduction}
%%%%%%%%%%%%%%%%%%%%%%%%%%%%
Black holes evaporate due to Hawking radiation \cite{Hawking:1975vcx}. Though this fact is on firm theoretical footing, there are still many open questions regarding the precise details of this evaporation, some of which may require a complete theory of quantum gravity to settle.

String theory provides such a theory. Besides strings, this framework also introduces other extended objects known as branes. These dynamical objects gravitate, and a large number of them can be described in the supergravity limit as a black brane. In analogy with Hawking radiation, one can ask if such black branes can be unstable to the emission of the individual branes from which they are formed. This has indeed been found to occur \cite{Yamada:2008em,Hartnoll:2009ns,Herzog:2009gd,Henriksson:2019zph,Henriksson:2019ifu}, for example by computing the effective potential for a brane probe and finding a global minimum \emph{outside} the horizon. In these cases the horizon and the true minimum are typically separated by a potential barrier; thus, the emission can be expected to proceed through bubble nucleation \cite{Coleman:1977py,Linde:1981zj}. Such \emph{brane nucleation} can provide the main mechanism for black branes to decay, and are thus important for a complete understanding of their quantum description. Moreover, through holographic duality \cite{Maldacena:1997re}, the brane nucleation can be interpreted as bubble nucleation in a dual field theory at strong coupling, a topic of interest in early-universe cosmology \cite{Hindmarsh:2020hop}.

In this Letter, we provide the first detailed study of a brane nucleation event. We consider a black brane, built from a stack of D3-branes at a conifold singularity, which emits a single D3-brane. We focus on two different aspects; the initial nucleation of the brane as a localized bubble stretching outward from the horizon, and the late-time steady-state expansion of this brane into the bulk. We do this by first deriving an effective action describing the embedding of the D3-brane in the black brane geometry. Then, we use this to find static configurations with $O(3)$ symmetry, whose Euclidean action gives the leading semi-classical contribution to the nucleation rate. Next, assuming a steady-state solution, we solve for the late-time expansion of the brane. In particular this lets us obtain the terminal velocity and profile of the brane. Throughout, we use the probe approximation, neglecting the backreaction of the nucleating D3-brane on the black brane background. 

The black branes we study live in an asymptotically anti--de Sitter (AdS) spacetime, and are holographically dual to states in the Klebanov-Witten (KW) gauge theory \cite{Klebanov:1998hh} at non-zero temperature and density. The D3-brane emission is then dual to the spontaneous breaking of the gauge group through the condensation of a scalar field dual to the radial position of the brane. Had the minimum of the brane's effective potential been at a finite radius (as is the case in some similar setups \cite{Henriksson:2019zph,Henriksson:2019ifu}) the nucleation would have resulted in a phase transition to a (metastable) Higgs phase, analogous to color superconductivity in QCD \cite{Rajagopal:2000wf}. For our D3-branes the minimum is instead located at the AdS boundary; this seems to indicate a fatal instability of the finite density KW theory, with the dual black brane slowly disappearing as one D-brane after the other nucleates and travels towards infinity (at least until the combined backreaction of the emitted branes can no longer be ignored).

%%%%%%%%%%%%%%%%%%%%%%%%%%%%
\section{Gravity solutions and field theory dual}
%%%%%%%%%%%%%%%%%%%%%%%%%%%%
We study asymptotically AdS$_5\times T^{1,1}$ black brane solutions of type IIB supergravity found by Herzog et al. \cite{Herzog:2009gd}. Defining forms with legs on $T^{1,1}$,
\begin{equation}
\begin{split}
\omega_2 &\equiv \frac{1}{2} \left(\sin \theta_1 d\theta_1 \wedge d\phi_1
- \sin \theta_2 d\theta_2 \wedge d\phi_2 \right) \\
g_5 &\equiv d \psi + \cos \theta_1 d\phi_1 + \cos \theta_2 d\phi_2 \ ,
\end{split}
\end{equation}
the ansatz for the 10D metric is
\begin{equation}\label{eq:10Dmetric}
\begin{split}
 ds_{10}^2 =& L^2 e^{-\frac{5}{3} \chi} ds_5^2 + L^2e^{\chi}
\bigg[\frac{e^{\eta}}{6} (d \theta_1^2 + \sin^2 \theta_1 d\phi_1^2)
\\
&+ \frac{e^{\eta}}{6} (d \theta_2^2 + \sin^2 \theta_2 d\phi_2^2)
+ \frac{e^{-4 \eta}}{9} g_5^2 \bigg] \ ,
\end{split}
\end{equation}
with $ds_5^2$ an asymptotically AdS$_5$ line element, and $L$ the corresponding AdS radius. Meanwhile, the self-dual 5-form field strength takes the form $F_5 = L^4 \left( {\cal F} + *{\cal F} \right)$, with
\begin{equation}\label{eq:5form}
{\cal F} = -\frac{2}{27} \omega_2 \wedge \omega_2 \wedge g_5 - \frac{1}{9\sqrt{2}} dA \wedge g_5 \wedge \omega_2 \ .
\end{equation}
Here $A$ is a $U(1)$ gauge field.

This ansatz provides a consistent truncation of type IIB supergravity to a 5D theory containing the metric, the gauge field $A$ and scalar fields $\chi$ and $\eta$. The resulting equations of motion have asymptotically AdS$_5$ charged black brane solutions; their 5D metric can be written as
\begin{equation}\label{eq:5Dmetric}
ds_5^2 = - g(r) e^{-2w(r)} dt^2 + \frac {dr^2}{g(r)} + r^2 d\vec x_3^2 \ ,
\end{equation}
with $g(r_H)=0$ at the horizon radius $r_H$. These solutions were constructed numerically and studied in \cite{Herzog:2009gd,Henriksson:2019ifu}, and we refer the interested reader there for more details.

String theory on asymptotically AdS$_5\times T^{1,1}$ is holographically dual to the KW theory \cite{Klebanov:1998hh}, and the supergravity limit we study describes it in the large-$N$, strong coupling limit. KW theory is an $\mathcal{N}=1$ superconformal theory with gauge group $SU(N)\times SU(N)$ and matter fields $A_\alpha$ and $B_{\tilde\alpha}$ ($\alpha,\tilde\alpha=1,2$) in the $(N,\bar N)$ and $(\bar N,N)$ representation, respectively. It provides the low-energy description of $N$ D3-branes placed on the tip of the cone with base $T^{1,1}$ (the conifold). Among the global symmetries a certain $U(1)$ factor is often referred to as ``baryonic'' since the only gauge invariant operators charged under it are heavy, having conformal dimensions of order $N$. The corresponding conserved current is dual to the $U(1)$ gauge field in the gravity truncation above. Thus, the black branes we study are dual to states of KW theory at finite temperature and baryonic chemical potential, computed by standard methods as
$T = e^{-w(r_H)}g'(r_H)/4\pi$ and $\mu = \Phi(r\to\infty)$.
Since the theory is conformal, the physics depends only on the dimensionless parameter $T/\mu$.

The lightest baryonic operators in KW theory involve determinants of the bifundamental matter fields, which are dual to D3-branes wrapped on a 3-cycle on $T^{1,1}$ \cite{Gubser:1998fp}. As the effective potential for these wrapped D3's does not show an instability \cite{Herzog:2009gd}, many of the usual ways a charged black brane can become unstable near extremality are ruled out. The only known instability is instead due to the emission of D3-branes parallel to the original stack, which is what we study in the following.

%%%%%%%%%%%%%%%%%%%%%%%%%%%%
\section{D3-brane effective action}
%%%%%%%%%%%%%%%%%%%%%%%%%%%%
We imagine that one of the D3-branes that build up the black brane localizes somewhere in the bulk. It then acts as a domain wall, which in the dual KW theory effectively changes the rank of the two $SU(N)$ gauge groups by one \cite{Gubser:1998fp}. Thus, if the energy can be lowered by separating a D3-brane from the black brane, the gauge symmetry in KW theory is spontaneously broken.

The action of a D3-brane is the sum of a Dirac-Born-Infeld (DBI) and a Wess-Zumino (WZ) term \cite{Johnson:2005mqa},
\begin{equation}\label{eq:D3action}
  S_{D3} = -T_3\int d^4\xi \sqrt{-\det P[G]}+T_3\int P[C_4] \ ,
\end{equation}
where $P[\dots]$ denotes the pullback of spacetime fields to the brane worldvolume. The constant dilaton has been absorbed into the tension of the brane, given by \cite{Herzog:2009gd}
\begin{equation}\label{eq:tension}
 T_3 = \frac{1}{(2\pi)^3 g_s l_s^4} = \frac{27}{32\pi^2}\frac{N}{L^4} \ .
\end{equation}

Letting the brane extend in the directions parallel to the horizon, we parameterize its worldvolume by the spacetime coordinates $\{t,\vec x\}$. The brane will be located on a constant and arbitrary point on $T^{1,1}$, while the embedding in the radial direction is taken to be a general function $R=R(t,\vec x)$. In the following, we denote derivatives with respect to $t$ by a dot and with respect to $x_i$ by $\partial_i$, $i=1,2,3$. The components of the 10D metric will be denoted by $G_{\mu\nu}$.

For such an embedding, the induced metric on the brane can be written as
\begin{equation}
 ds^2_4 = \gamma_{tt} dt^2 + 2\gamma_{ti} dt\, dx^i + \gamma_{ij} dx^i dx^j \ ,
\end{equation}
where the components of the induced metric are given by
\begin{equation}
\begin{split}
 \gamma_{tt} &= G_{tt} + G_{rr} \dot R^2 \\
 \gamma_{ti} &= G_{rr}\dot R\, (\partial_i R) \\
 \gamma_{ij} &= G_{ij} + G_{rr}(\partial_i R)(\partial_j R) \ .
\end{split}
\end{equation}
The determinant in the DBI term is then
\begin{equation}
 \det P[G] = G_{tt}G_{xx}^3\left( 1+\frac{G_{rr}}{G_{tt}}\dot R^2 + \frac{G_{rr}}{G_{xx}} (\partial_i R)^2 \right) \ .
\end{equation}
For the WZ term we need the pullback of the 4-form potential defined by $F_5=dC_4$, the relevant component of which is $L^4 a_4(r)\, dt\wedge dx_1\wedge dx_2\wedge dx_3$,
where $a_4(r)$ is found by integrating
$a_4'(r) = 4r^3e^{-w-\frac{20}{3}\chi}$
with the condition that it goes to zero on the horizon. Plugging in the metric components and using (\ref{eq:tension}), the full action becomes
\begin{equation}\label{eq:fullAction}
\begin{split}
 S_{D3} &= -\frac{27N}{32\pi^2} \int d^4x\, \Bigg\{ R^3 e^{-w(R)-\frac{10}{3}\chi(R)}\\
 &\sqrt{ g(R) - \frac{e^{2w(R)}}{g(R)}\dot R^2 + \frac{(\partial_i R)^2}{R^2}  } - a_4(R) \Bigg\} \ .
\end{split}
\end{equation}

In the dual KW theory, the radial position of the brane maps to a combination of eigenvalues of the matrices $A_\alpha$ and $B_{\tilde\alpha}$. If the energy can be lowered by a brane separating from the stack, the dual field will condense, causing the gauge symmetry breaking $SU(N)\times SU(N)\to SU(N-1)\times SU(N-1)\times U(1)$. We can then interpret (\ref{eq:fullAction}) as a quantum effective action describing the dynamics of the scalar field in the resulting $U(1)$ sector.

The effective potential is obtained by setting all derivatives of $R$ in (\ref{eq:fullAction}) to zero. Evaluating it on the backgrounds discussed above, one finds an instability --- a global minimum at $r=\infty$ --- for $T/\mu\lesssim 0.2$ \cite{Herzog:2009gd,Henriksson:2019ifu}. We now proceed to study the dynamics of this instability.

%%%%%%%%%%%%%%%%%%%%%%%%%%%%
\section{Bubble nucleation}\label{sec:bubbles}
%%%%%%%%%%%%%%%%%%%%%%%%%%%%
To compute the nucleation rate of the transition, we analytically continue to Euclidean time and take the time direction to have periodicity $1/T$. We search for static solutions with a spherical $O(3)$ symmetry --- such solutions give an accurate estimate for the nucleation rate at sufficiently high temperatures, and provide an upper bound at low temperatures. The $t$-integral can be done right away, giving a factor $1/T$. Due to the anticipated $O(3)$ symmetry we switch to spherical coordinates $\{\rho,\alpha,\beta\}$ on the worldvolume, take $R=R(\rho)$, and integrate over $\alpha$ and $\beta$ as well. This gives us the Euclidean action
\begin{equation}
 S_{D3} = \frac{27N}{8\pi T} \int d\rho\, \rho^2 \left\{ \frac{R^2\sqrt{ g R^2 + (R')^2 }}{e^{w+\frac{10}{3}\chi}} - a_4 \right\} \ , \label{eq:D3actionEuclidean}
\end{equation}
where a prime denotes a $\rho$-derivative. Here and below we suppress the radial dependence of $g$, $w$, $\chi$ and $a_4$. Given a solution to the equation of motion (EoM) resulting from this action, the nucleation rate per volume is estimated as $\Gamma/V \sim e^{-S_{D3}}$ \cite{Coleman:1977py,Linde:1981zj}.

\begin{figure}[t!]
 \centering
 \includegraphics[width=8.6cm]{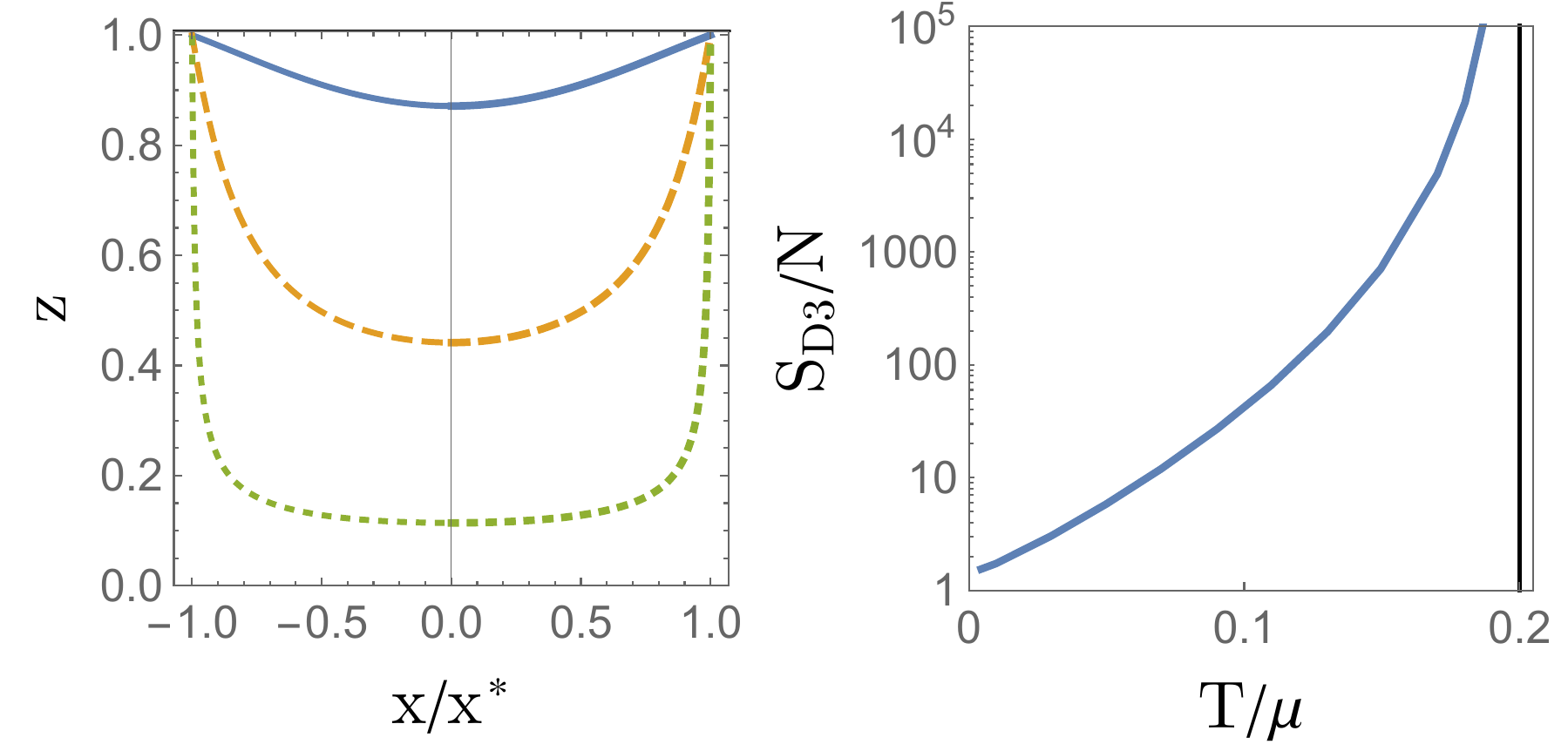}
 \caption{Left: Bubble cross sections at $T/\mu=0.01$ (solid blue), $T/\mu=0.09$ (dashed orange), and $T/\mu=0.18$ (dotted green), using the radial coordinate $z=r_H/r$ and with the field theory direction $x$ rescaled in each case such that the bubble radius is 1. Right: The action evaluated on the bubble solutions as a function of $T/\mu$.}
 \label{fig:bubbles}
\end{figure}

The EoM is a second-order ordinary differential equation. Solutions should satisfy $R'(0)=0$ to be smooth at the origin. We can find a series solution near $\rho=0$ with $R(0)$ a free parameter. We furthermore expect our solutions to hit the horizon located at $R=r_H$ at some finite $\rho=\rho^*$. We can then also find a series solution near this point of the form
\begin{equation}
 R(\rho) = r_H + R_1(\rho^*-\rho) + R_2(\rho^*-\rho)^2 + \cdots \ ,
\end{equation}
where everything except the value of $\rho^*$ is fixed by the EoM. Using these two series expansions, we construct the full solutions by a two-sided numerical shooting method, fixing the parameters $R(0)$ and $\rho^*$ by matching the solutions and their first derivative at some intermediate radius. The resulting bubble cross sections at three different $T/\mu$ are shown on the left in Fig.~\ref{fig:bubbles}.

We then evaluate the action (\ref{eq:D3actionEuclidean}) on these bubble solutions. Note that it diverges in the large-$N$ limit, suppressing the nucleation rate. This is to be expected as the large-$N$ limit is a classical limit on the gravity side. On the right in Fig.~\ref{fig:bubbles} we show the action divided by $N$ evaluated on our bubble solutions as a function of $T/\mu$. Approaching the critical value $T/\mu\approx 0.2$ it diverges (even for finite $N$), as expected on general grounds. At low temperatures, it approaches an $\OO(1)$ number, implying that nucleation is still greatly suppressed for $N$ large. Recall however that at low temperatures our static solutions in general only provide an upper bound on the action, and one should also consider solutions with dependence on Euclidean time. We leave this for future work.

Note that the static bubble solutions discussed so far can be ``completed'' in two distinct ways, due to the fact that a brane laying flat on the horizon is static in the relevant time coordinate (because of the infinite time dilation). One way is to add such a horizon brane stretching from the edge of the bubble out to infinity in the $\vec x$ directions. The other way is to add a horizon \emph{antibrane} inside the bubble. We interpret the former completion as the emission of a D-brane from the horizon, and the latter as brane-antibrane pair creation with the antibrane falling into the horizon and the brane escaping to infinity. Since an (anti)brane on the horizon has zero action, neither of these completions affect the nucleation rate and both are equally likely to occur.

%%%%%%%%%%%%%%%%%%%%%%%%%%%%
\section{Late-time expansion}\label{sec:wallSpeed}
%%%%%%%%%%%%%%%%%%%%%%%%%%%%
As a bubble nucleates, a small perturbation might cause it to start expanding. At late times, the resulting bubble wall can be approximated as planar, moving in (say) the $x_1$-direction. Then it is useful to parameterize the worldvolume by its $r$-coordinate instead of its $x_1$-coordinate, and work with the embedding function $X_1\equiv X(t,r)$. This switch results in the action
\begin{equation}\label{eq:actionX}
\begin{split}
 S_{D3} = &-\frac{27N}{32\pi^2} \int dt\, dr\, dx_2\, dx_3 \Bigg\{ r^2 e^{-w-\frac{10}{3}\chi} \\
 &\sqrt{ 1 - \frac{e^{2w} r^2}{g}\dot X^2 + r^2 g\, X'^2 } + X'a_4 \Bigg\} \ ,
\end{split}
\end{equation}
where a prime now denotes an $r$-derivative. Since only derivatives of $X$ appear in the action, there is an associated conserved current describing the brane's momentum density, and the EoM is just the conservation of this current.

At late times we expect the bubble wall to reach a terminal velocity, with outward pressure being balanced by friction. Thus we search for steady-state solutions of the form
\begin{equation}
 X(t,r)=vt + \xi(r) \ .
\end{equation}
The calculation now proceeds much like the classical drag force calculations \cite{Herzog:2006gh,Gubser:2006bz}; see also \cite{Bigazzi:2021ucw}. The ansatz leads to the simplified EoM $\partial_r P^r=0$, with the radial component of the momentum current being
\begin{equation}
 P^r = -\frac{r^4 g\, e^{-w-\frac{10}{3}\chi}}{ \sqrt{ 1 - \frac{e^{2w}}{g}r^2v^2 + r^2 g\, \xi'(r)^2 } }\, \xi'(r) - a_4 \ .
\end{equation}
Solving for $\xi'(r)$ gives
\begin{equation}\label{eq:xiPrime}
 \xi'(r) = \pm \frac{P^r + a_4}{g}\sqrt{\frac{e^{2w}v^2-g/r^2}{\left[P^r + a_4\right]^2 - r^6 e^{-2w-\frac{20}{3}\chi} g}} \ .
\end{equation}
Before attempting to integrate this equation, let us discuss what we expect our solution to look like. As a brane nucleates and then starts to expand, it gains energy and momentum from the region near its center, which is at a larger radius where the potential energy is lower. This accelerates the brane, while some energy is dissipated into the near-horizon region --- this is the gravity dual of the friction between the bubble and the plasma on the field theory side. The end result at late times should be a solution extending from the horizon towards the true minimum at infinity. Thus, we must impose that $\xi'(r)$ diverges as $r\to\infty$. This can only happen if the denominator in the square root of (\ref{eq:xiPrime}) goes to zero there, fixing $P^r$ to be
\begin{equation}
 P^r = \lim_{r \to +\infty} -a_4 \pm r^3 e^{-w-\frac{10}{3}\chi} \sqrt{g} \ ,
\end{equation}
where we must pick the plus sign to get a finite result. This sets $P^r$ equal to the minimum of the effective potential.

\begin{figure}[t!]
 \centering
 \includegraphics[width=8.6cm]{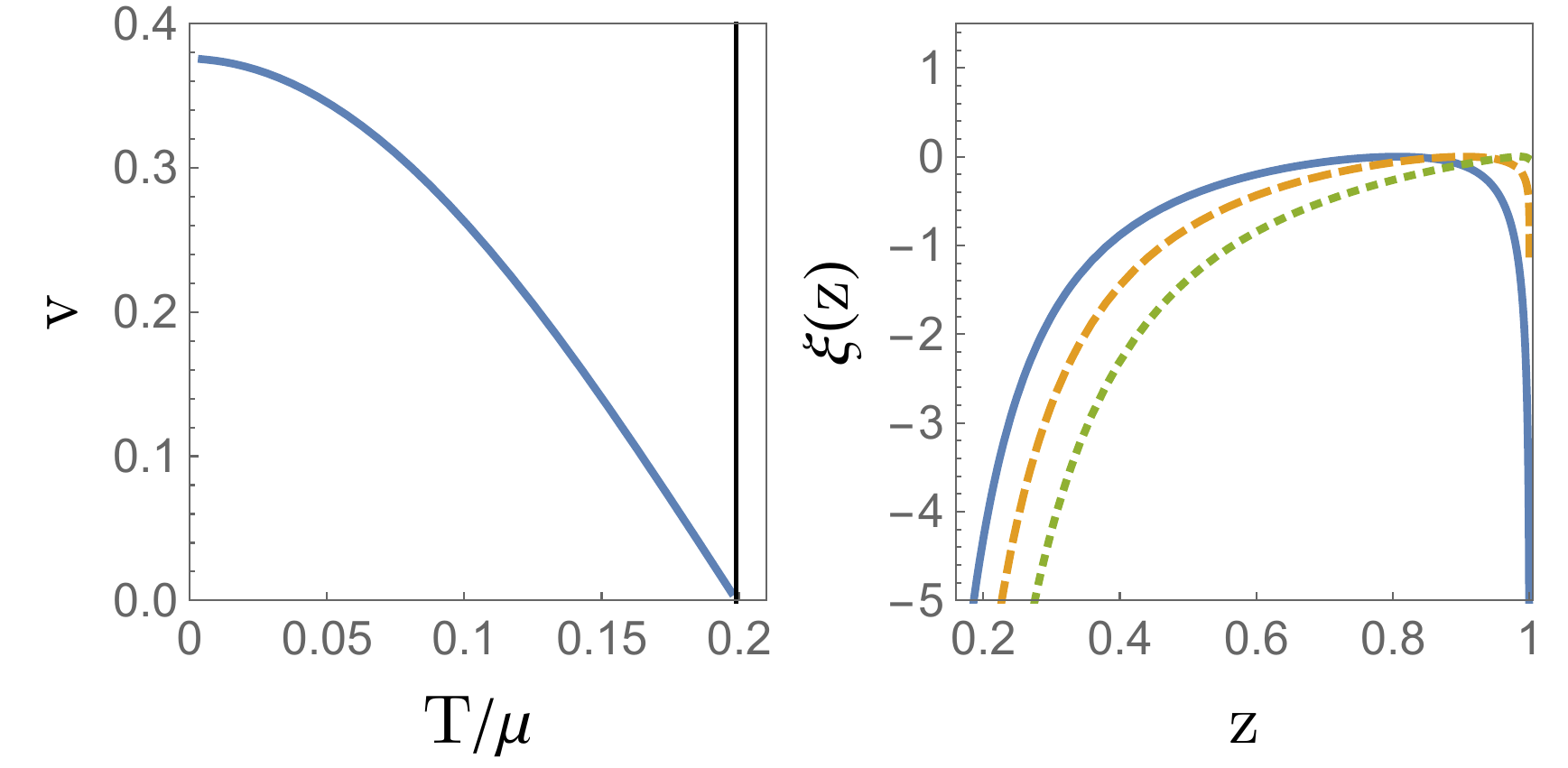}
 \caption{Left: The terminal wall speed as a function of $T/\mu$. Right: The wall profile as a function of the radial coordinate $z=r_H/r$ at $T/\mu=0.01$ (solid blue), $T/\mu=0.09$ (dashed orange), and $T/\mu=0.18$ (dotted green). Note that in all cases the curves diverge to infinity at the horizon, though this is not visible for the green curve.}
 \label{fig:wallSpeedAndProfile}
\end{figure}

Having fixed $P^r$ thus, we notice that the denominator inside the square root of (\ref{eq:xiPrime}), in addition to going to zero at infinity, also crosses zero at some finite $r$. This forces us to fix $v$ such that the numerator crosses zero at the same point, ensuring that $\xi'(r)$ is everywhere real. Thus we arrive at a numerical value for the wall speed of the expanding bubble at asymptotic times, shown as a function of $T/\mu$ on the left in Fig.~\ref{fig:wallSpeedAndProfile}. As expected, the speed goes to zero at the critical temperature. For a relativistic theory at zero density the speed would approach the speed of light at small temperatures; here however, it approaches a smaller value of about 0.38. Note that the wall speed remains well below the speed of sound of the background plasma, which is fixed to the conformal value of $1/\sqrt{3}$. In the usual parlance of first order phase transitions this would thus be classified as a deflagration (as opposed to a detonation), although in the probe approximation this is a somewhat artificial distinction since the bubble doesn't backreact on the plasma.

With the parameters fixed, (\ref{eq:xiPrime}) can be numerically integrated to find the profile $\xi(r)$ of the moving wall. Assuming $v>0$, $\xi(r)$ must diverge to negative infinity at the AdS boundary. This fixes the sign ambiguity in (\ref{eq:xiPrime}), but only partially, since the derivative $\xi'(r)$ always goes to zero at some point in the bulk. (This can be seen from (\ref{eq:xiPrime}), recalling that $P^r$ is negative and $a_4(r)$ increases monotonically from zero at the horizon.) At this point we are free to change the sign used in (\ref{eq:xiPrime}) without creating a discontinuity in the derivative. Thus, there are two possible late-time wall profiles; we fix this final ambiguity by requiring that the part of the brane near the horizon trails behind the front of the wall, much like in \cite{Herzog:2006gh,Gubser:2006bz}. The resulting wall profile is shown on the right of Fig.~\ref{fig:wallSpeedAndProfile}. Note that $\xi(r)$ diverges to negative infinity also at the horizon, as can be seen by expanding (\ref{eq:xiPrime}) there. 

%%%%%%%%%%%%%%%%%%%%%%%%%%%%
\section{Discussion}\label{sec:discussion}
%%%%%%%%%%%%%%%%%%%%%%%%%%%%
For the charged black branes studied in this Letter, brane nucleation appears to be the leading channel of decay. One might thus expect the resulting evaporation process to lead to an information paradox in much the same way as with the more familiar Hawking radiation \cite{Hawking:1976ra}. In fact, brane nucleation offers a way of studying the disappearance of a large black brane (or a black hole \cite{Henriksson:2019zph}) in AdS without modifications at the boundary \cite{Rocha:2008fe}. This can lead to new ideas for solving the information paradox, and new ways of testing such ideas.

From the point of view of the dual field theory, we have studied barrier penetration through bubble nucleation at strong coupling. While in this case the nucleation does not appear to result a new (meta)stable phase in the field theory, our methods extend straightforwardly to other setups where this would be the case (since the minimum of the effective potential is at a finite radius) \cite{Henriksson:2019zph,Henriksson:2019ifu}. There, brane nucleation mediates a genuine first order phase transition. Much recent work has leveraged holographic duality to study bubble nucleation \cite{Bigazzi:2020phm,Agashe:2020lfz,Bigazzi:2020avc,Ares:2020lbt,Bea:2021zsu,Bigazzi:2021ucw}, motivated by early-universe phase transitions and the gravitational waves they could produce \cite{Hindmarsh:2020hop}. This Letter is to our knowledge the first where both nucleation rates and the bubble wall speed have been computed from first principles, using simple numerical tools and without any additional approximations beyond treating the nucleating brane as a probe.

When a metastable endstate exists, another interesting quantity which can readily be computed is the surface tension between two coexisting phases. Furthermore, one could use the effective action (\ref{eq:fullAction}) to study soliton solutions and Wilson line configurations along the lines of \cite{Schwarz:2014rxa,Evans:2019pcs,Kumar:2020hif}. We also hope to study brane solutions with general time-dependence, which would give a more complete picture of the expansion of the bubble and, in the case of Euclidean time, a better estimate of the nucleation rate at low temperatures. Going away from the high-temperature limit might also lead to a clearer separation between the two possible completions of the static bubble solutions, which we have interpreted as brane emission and brane-antibrane pair creation, respectively.

Lastly, it would be very interesting to include backreaction from the bulk D-branes on the geometry, at least in some approximate way, as this is the only way to search for the true end state of a brane nucleation instability.

%%%%%%%%%%%%%%%%%%%%%%%%%%%%

\begin{acknowledgments}
\inlinesection{Acknowledgments}
It is a pleasure to thank Francesco Bigazzi, Aldo Cotrone and Carlos Hoyos for helpful comments and suggestions, and F{\"e}anor Reuben Ares, Mark Hindmarsh, Niko Jokela, Otto Karhu, Huaiyu Li and Niilo Nurminen for useful discussions and collaboration on related work. The work was supported by the Academy of Finland (grant number 330346), the Ruth and Nils-Erik Stenbäck foundation, and the Waldemar von Frenckell foundation.
\end{acknowledgments}

%%%%%%%%%%%%%%%%%%%%%%%%%%%%

\bibliographystyle{apsrev4-2}

\bibliography{biblio}

\end{document}